# On-Demand Generation of Neutral and Negatively-Charged Silicon-Vacancy Centers in Diamond

Siddharth Dhomkar[1,*], Pablo R. Zangara[1,*], Jacob Henshaw[1,2,*], and Carlos A. Meriles[1,2,†]

[1]*Dept. of Physics, CUNY-City College of New York, New York, NY 10031, USA.*
[2]*CUNY-Graduate Center, New York, NY 10016, USA.*

Point defects in wide-bandgap semiconductors are emerging as versatile resources for nanoscale sensing and quantum information science but our understanding of the photo-ionization dynamics is presently incomplete. Here we use two-color confocal microscopy to investigate the dynamics of charge in Type 1b diamond hosting nitrogen-vacancy (NV) and silicon-vacancy (SiV) centers. By examining the non-local fluorescence patterns emerging from local laser excitation, we show that in the simultaneous presence of photo-generated electrons and holes, SiV (NV) centers selectively transform into the negative (neutral) charge state. Unlike NVs, 532 nm illumination ionizes SiV$^-$ via a single photon process thus hinting at a comparatively shallower ground state. In particular, slower ionization rates at longer wavelengths suggest the latter lies approximately ~1.9 eV below the conduction band minimum. Building on the above observations we demonstrate on-demand SiV and NV charge initialization over large areas via green laser illumination of variable intensity.

An intense effort is underway to understand and exploit the properties of select paramagnetic centers in solids, with the nitrogen vacancy (NV) and silicon-vacancy (SiV) centers in diamond arguably standing out as the most prominent examples[1,2]. Recent work on the photo-physics of the NV has shown that the defect's charge state is dynamically modulated between neutral and negative configurations during green illumination[3,4]. Red excitation, on the other hand, can ionize the negatively charged NV (NV$^-$) but leaves the neutral state (NV$^0$) unchanged. These and related processes have been exploited, e.g., to demonstrate alternate forms of NV spin sensing, including photo-electric detection of magnetic resonance[5] and spin-to-charge conversion[6]. Spin selective charging dynamics of NV and N centers is also of interest to recent proposals of long-range spin buses for diamond quantum computing[7].

Similar to the NV, SiV centers are known to exist in neutral (SiV$^0$) and negatively charged (SiV$^-$) states but a thorough understanding of charge conversion is still lacking[8]. The negatively charged state is believed to be more stable in Type 1b diamond but efforts have been made to tilt the balance, for example, by resorting to boron doping[9]. Here we show that SiV$^0$ efficiently captures photo-generated electrons diffusing from a remote region of the diamond crystal; once in the negative state, however, SiVs are immune to photo-generated holes. This behavior mirrors that of NVs, readily transitioning from the negative to the neutral state through hole capture, though stable in the presence of conduction-band electrons. By probing the time dependent response of SiV$^-$ to green illumination we find a linear growth of the ionization rate with laser power, the hallmark of a single-photon process.

In our experiments we use a [111] crystal with an estimated N content $P$ between 1 and 5 ppm, and rely on a custom-made, two-color confocal microscope to excite and probe NV and SiV fluorescence over a ~1-μm-diameter spot[10,11]. Optical spectroscopy is carried out with the help of a high-resolution optical spectrometer. A typical spectrum under green (532 nm) excitation is shown in Fig. 1a: Given the characteristic zero-phonon lines (ZPL) at 575 nm and 637 nm we conclude the fluorescence largely originates from NV centers in the neutral and negative states, respectively. The small feature at 737 nm reveals the presence of SiV$^-$ centers, though given the low fractional intensity — ~0.6% of the emitted photons — it seems reasonable to anticipate a marginal impact on the overall system dynamics. Unexpectedly, however, the experiments below show that the relative SiV$^-$ population — and consequently the contribution of SiV$^-$ fluorescence to the integrated optical signal — can change dramatically depending on the experimental conditions and illumination history.

The detection protocol in Fig. 1b provides an initial demonstration: We first use multiple scans of a 632 nm laser beam to initialize NVs into the neutral charge state over a (40 μm)$^2$ area. Since red light also ionizes SiV$^-$ (see below), this procedure helps us circumvent the more complex dynamics during green illumination (which we discuss later). We subsequently park the red beam at the center point so as to ionize neutral nitrogen. The process demands a multi-second long exposure because N$^0$ absorbs weakly at 632 nm[18-21]. In this sample, however, the nitrogen concentration is sufficient to produce a substantive number of photo-generated electrons, some of which diffuse away from the illuminated area to be subsequently captured by other proximal defects[10]. To reveal the ensuing charge distribution we acquire a confocal image of the area of interest using a weak red laser scan. The results using NV$^-$-selective and SiV$^-$-selective band-pass filters (respectively, BP2 and BP3 in Fig. 1a) are presented in Fig. 1c. Interestingly, SiVs close to the point of prolonged beam exposure readily transform to the bright, negatively-charged state while neutral NVs remain unaffected. Note that these observations correct and extend prior work[10], where the role of SiV centers was overlooked.

To attain representative spectra before and after prolonged illumination, we average the signal resulting from a red laser scan across a 10×10 μm$^2$ area centered at the point of the laser park (respectively orange and blue traces in Fig. 1d). The prominent peak at 737 nm emerging after the scan confirms the selective formation of SiV$^-$ with virtually no impact on the original NV$^-$ concentration.

---

[†]E-mail: cmeriles@ccny.cuny.edu.

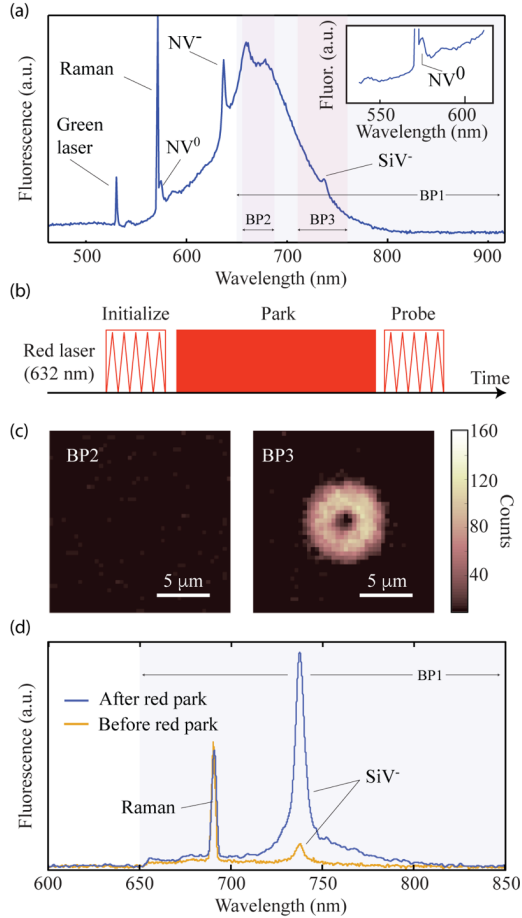

**Fig. 1:** (a) Optical spectrum under 532 nm excitation at an arbitrary but fixed location. The laser power is 100 μW and the acquisition time is 120 s. Color bands – denoted as BP1, BP2, and BP3 – indicate three alternative observation windows. (b) Initialization and optical spectroscopy protocol. The zig-zag indicates laser scanning. (c) Confocal images in an area around the point of laser park using bandpass filters BP2 (left) and BP3 (right). The park time is 30 s, the laser power is 1 mW during the park and 100 μW during imaging at 2 ms integration time per pixel. (d) Averaged optical spectra (632 nm excitation) from an area around the point of laser parking before and after the park (orange and blue traces, respectively). The red laser power during the spectroscopy scan is 500 μW; the acquisition time is 5 ms per pixel.

The ability to initialize SiV into the negative state gives us the opportunity to investigate the mechanisms underlying SiV$^-$ ionization. Fig. 2a shows the SiV$^-$ response when exposed to a 532 nm laser pulse of variable duration and power expressed in terms of the relative SiV$^-$ population $S_-/S$, which we determine via a calibrated red readout[11]; here we use $S_-$ ($S_0$) to denote the concentration of negative (neutral) SiV, and $S \equiv S_- + S_0$ is the total SiV concentration. On a time scale of seconds, we observe an exponential decay at a rate proportional to the laser power (green dots in Fig. 2b), indicative of ionization via a one-photon process. The ionization rate also grows linearly with intensity under 632 nm illumination[11], though the absolute values are much lower (the dashed line in Fig. 2b shows the extrapolated linear fit). We observe virtually no ionization at or above 690 nm, suggesting that the energy gap separating the SiV$^-$ ground state from the conduction band minimum amounts to ~1.9 eV. This value agrees reasonably well with prior estimates derived indirectly from photo-ionization experiments[22,23] or from density functional calculations[24], though is in conflict with more recent work[25,26]. In particular ab-initio calculations[25] sets the SiV$^-$ ground state ~0.8 eV above the valence band implying that the single-photon process observed in Fig. 2 corresponds, in reality, to a transformation into SiV$^{2-}$ via emission of a hole. The latter, of course, contradicts the interpretation above, namely, that red illumination ionizes SiV$^-$ into SiV$^0$, and that the back transformation takes place through the capture of electrons injected during the red park.

Since the ZPL emission from SiV$^0$ at 946 nm is difficult to observe directly at room temperature[8,9], the problem above implies ambiguity in the SiV charge state following the initialization scan. To reveal the type of carrier involved in the capture process we use metal electrodes to produce an electric field parallel to the direction of illumination (Fig. 2c). As the voltage grows, we see greater (lower) SiV$^-$ formation near the positive (negative) electrode (Figs. 2d-2f), indicating that free electrons, not holes, form during the red laser park. This observation strongly suggests that SiVs must be neutral prior to carrier capture. Further work will, therefore, be necessary to reconcile experiment and theory.

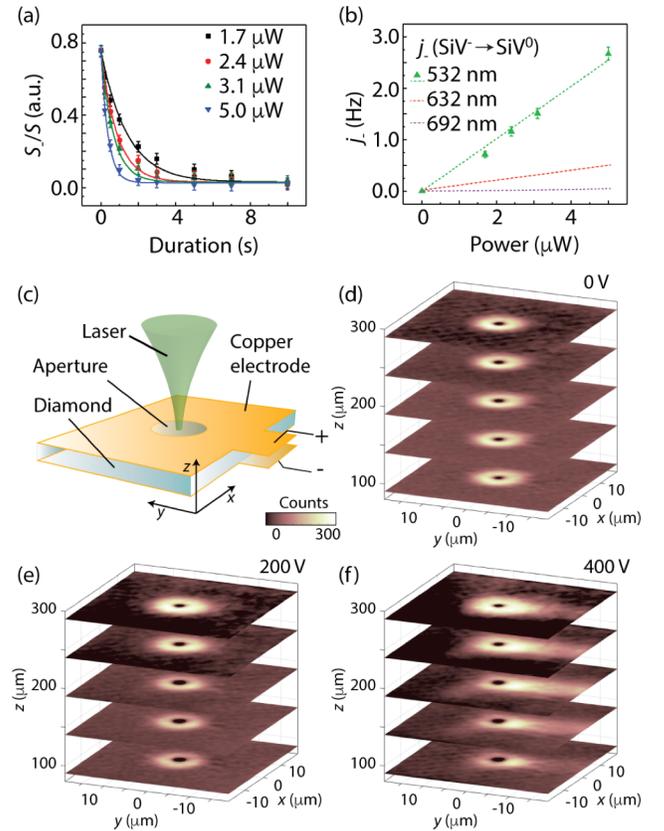

**Fig. 2:** (a) Relative population of SiV$^-$ ($S_-/S$) as a function of time for 532 nm laser light of variable power; readout is carried out via a red laser probe; solid lines are single exponential fits. (b) SiV$^-$ ionization rate ($j_-$) as a function of laser power for 532 nm laser excitation (green triangles). The dashed green line is a linear fit. Also shown are the extrapolated fits from SiV$^-$ ionization at 632 and 692 nm (red and purpled dashed lines, respectively). (c) Electrode and illumination geometry. (d-f) SiV$^-$-selective images after a 30-s-long red laser park at different depths. From (d) through (f) the applied voltage is 0V, 200V, and 400V, respectively. The distance from the top (bottom) image plane to the top (bottom) diamond surface is 10 (90) μm. All images share the same color scale and are collected during a red laser scan (1 mW, 2 ms integration time per pixel); all other conditions as in Fig. 1c.

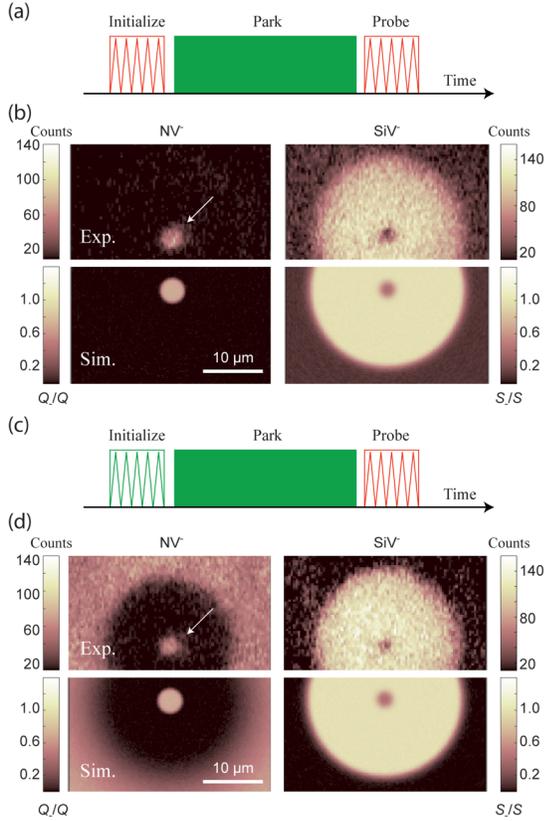

**Fig. 3:** (a) Pulse sequence. Red zig-zags indicate 632 nm, 1 mW laser scan and the green rectangle denotes a 532 nm, 1 mW laser park. (b) Confocal images (top) of the NV$^-$ (left) and SiV$^-$ (right) fluorescence after a 10 s green laser park (1 mW) obtained using the BP2 and BP3 bandpass filters. The calculated NV$^-$ and SiV$^-$ relative populations (respectively, $Q_-/Q$ and $S_-/S$) are shown in the bottom half images. (c,d) Same as above but after initialization using a 532 nm, 50 μW laser scan. In (b) and (d) the fluorescence within BP3 has been corrected to eliminate contributions from NVs. In all images the red laser power is 100 μW and the integration time per pixel is 2 ms; the white arrow indicates the point of laser parking.

We now examine the dynamics of carrier diffusion and trapping under green excitation, more complex due to the simultaneous diffusion of photo-generated electrons and holes. We start with the protocol in Fig. 3a, where we park the green beam in an NV$^-$-depleted region. Confocal images selective to NV$^-$ or SiV$^-$ fluorescence from an area around the point of laser parking are presented in Fig. 3b (upper left and right, respectively). Direct light exposure partly transforms NV$^0$ into NV$^-$ while simultaneously ionizing SiV$^-$ (bright spot and dark inner disk in the upper left and right images of Fig. 3b, respectively). Similar to Fig. 1c (red park), we observe a bright SiV$^-$ halo around the point of illumination, reflective of efficient electron capture. Interestingly, we find that the SiV$^-$ fluorescence intensities across the patterns in Figs. 1c and 3b show comparable amplitudes, indicating that the SiV$^-$ population remains unchanged in both cases. Given the efficient production of holes from green-laser-induced NV$^0$→NV$^-$ conversion, we conclude that the SiV$^-$ hole capture cross section is negligible. Further, since the charge distribution is uniform throughout the bright torus, we surmise SiVs saturate to collectively adopt the negatively charged state; below we use the observed fluorescence as a reference standard of SiV$^-$ fractional population.

To more clearly expose the role of holes in the above experiments we run a modified version of the protocol in Fig. 3a using a green scan for background initialization (Fig. 3c); the laser power is chosen so as to attain an NV$^-$-rich, SiV$^-$-depleted composition (see below). Comparison between Figs. 3b and 3d shows little change for SiVs, but the response is different for NVs where a large NV$^-$-depleted halo forms around the point of green laser parking, indicative of hole capture. These results expose an interesting correspondence, namely, the charge state of SiVs is immune to diffusing holes, whereas the charge state of NVs is stable in the presence of free electrons.

The interplay between carrier photo-generation, diffusion, and trapping can be formally described using a set of master equations for the populations of each defect species and equilibrium concentrations of electrons and holes[10,11]. Using parameters from the literature whenever possible, we calculate the distribution of trapped charge after a given interval of local green laser excitation in Figs. 3b and 3d for NV$^-$-depleted and NV$^-$-rich initial backgrounds, respectively (the starting SiV$^-$ content is low in both cases). Comparison with experiment yields reasonable agreement[11], though the calculated transition between bright and dark areas tends to be sharper (SiV$^-$) or smoother (NV$^-$) than observed.

The non-local impact of local laser excitation can be exploited to create large areas with controllable SiV$^-$ content. The idea is sketched in Fig. 4a for the case of a green beam scanning a uniform distribution of SiVs: Color centers exposed to the beam transform to (or remain in) the neutral state, while those in their vicinity become negative (respectively, areas *I* and *II* in the insert to Fig. 4a). The average charge distribution emerging from a scan, however, differs radically from the static pattern because, as the beam moves from one point to the next, any SiV$^0$ the beam produces locally is subsequently transformed into SiV$^-$ remotely via electron capture. The cumulative effect, therefore, is a controllable SiV$^-$ population, increasing with the laser brightness. Note that the converse argument applies to the co-existing ensemble of NVs, the majority of which must adopt a neutral state through the capture of holes photo-generated non-locally. This latter scenario may seem counterintuitive, as NV$^-$ initialization through green light exposure is the standard charge preparation protocol.

The results of our experiments — comprising a green initialization scan followed by a weak red readout scan — are presented in Figs. 4b and 4c: We first work in the limit of 'weak' laser excitation (Fig. 4b), where the effects of carrier diffusion away from the area under illumination can be neglected. Comparing the NV$^-$ fluorescence throughout the scanned area with that obtained after a green laser park, we conclude that the fractional NV$^-$ population amounts to ~70%, close to the maximum possible. By contrast, the SiV$^-$ fractional population remains at background levels, a consequence of the local ionization by 532 nm illumination. Alternatively, when the green beam is strong enough to generate a large number of carriers, SiVs adopt the negatively charged state whereas NVs are predominantly initialized into the neutral state (Fig. 4c).

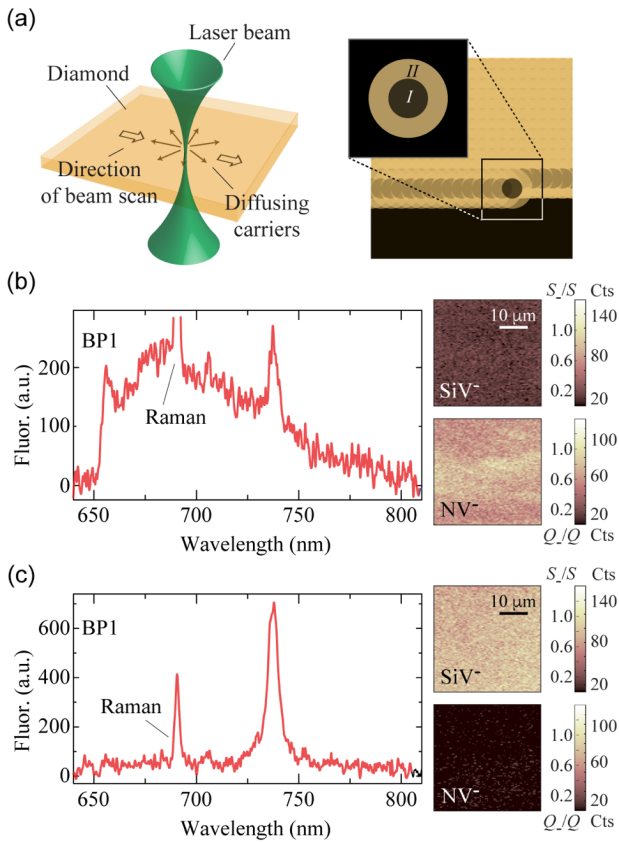

**Fig. 4:** (a) (Left) As the green beam scans the diamond crystal, electrons and holes diffuse away from the illuminated spot. (Right) Effect of a strong green scan on SiV$^-$; *I* and *II* in the insert indicate the areas exposed to the beam and affected by electron capture, respectively. (b) (Right) SiV$^-$-selective (top) and NV$^-$-selective (bottom) scanning confocal images (632 nm readout) obtained after a 532 nm, 50 μW scan. (Left) Optical spectrum averaged over the same area using bandpass BP1. (c) Same as in (b) but for a 1 mW green laser initialization scan. The integration time during the confocal imaging and spectroscopy scans are 2 ms and 5 ms, respectively; in both cases the red laser power is 100 μW. All other conditions as in Fig. 1.

In summary, SiV$^-$ ionizes via a one-photon process when excited with green or red light; negligible ionization rates at or above 690 nm suggest the energy gap separating the SiV$^-$ ground state from the conduction band minimum is approximately 1.9 eV. Further work will be required to reconcile these observations and the present ab-initio model of SiV$^-$.[25] By observing the distribution of trapped charge after local photo-ionization we find that neutral SiVs capture electrons efficiently from the conduction band; the negative-charge state remains stable over time, even in the presence of diffusing holes. By contrast, NVs are seen to be effective hole traps, immune to conduction electrons. In all cases, the defect charge state is stable in the dark (we detect no change on a time span of a week) consistent with prior work.[10,27,28] Looking forward, our results should prove valuable for the various ongoing experiments aimed at harnessing the properties of the SiV, either as a non-linear optical switch[29], or as an optically addressable spin qubit[9,30]. These findings could also find application, for example, to enhance the storage capacity and number of non-destructive charge readouts in diamond-based memories.[28]

We thank Marcus Doherty, Adam Gali, and Harishankar Jayakumar, for fruitful discussions and assistance with some of the experiments. All authors acknowledge support from the National Science Foundation through grants NSF-1619896, NSF-1547830, and from Research Corporation for Science Advancement through a FRED Award.

**References**

[1] M.W. Doherty, N.B. Manson, P. Delaney, F. Jelezko, J. Wrachtrup, L.C.L. Hollenberg, *Phys. Rep.* **528**, 1 (2013).

[2] L.J. Rogers, K.D. Jahnke, M.H. Metsch, A. Sipahigil, J.M. Binder, T. Teraji, H. Sumiya, J. Isoya, M.D. Lukin, P. Hemmer, F. Jelezko, *Phys. Rev. Lett.* **113**, 263602 (2014).

[3] G. Waldherr, J. Beck, M. Steiner, P. Neumann, A. Gali, Th. Frauenheim, F. Jelezko, J. Wrachtrup, *Phys. Rev. Lett.* **106**, 157601 (2011).

[4] N. Aslam, G. Waldherr, P. Neumann, F. Jelezko, J. Wrachtrup, *New J. Phys.* **15**, 013064 (2013).

[5] E. Bourgeois, A. Jarmola, P. Siyushev, M. Gulka, J. Hruby, F. Jelezko, D. Budker, M. Nesladek, *Nat. Commun.* **6**, 8577 (2015).

[6] B.J. Shields, Q.P. Unterreithmeier, N.P. de Leon, H. Park, M.D. Lukin, *Phys. Rev. Lett.* **114**, 136402 (2015).

[7] M.W. Doherty, C.A. Meriles, A. Alkauskas, H. Fedder, M.J. Sellars, N.B. Manson, *Phys. Rev. X* **6**, 041035 (2016).

[8] U.F.S. D'Haenens-Johansson, A.M. Edmonds, B.L. Green, M.E. Newton, G. Davies, P.M. Martineau, R.U.A. Khan, D.J. Twitchen, *Phys. Rev. B* **84**, 245208 (2011).

[9] B.C. Rose, D. Huang, Z-H. Zhang, A.M. Tyryshkin, S. Sangtawesin, S. Srinivasan, L. Loudin, M.L. Markham, A.M. Edmonds, D.J. Twitchen, S.A. Lyon, N.P. de Leon, arXiv:1706.01555.

[10] H. Jayakumar, J. Henshaw, S. Dhomkar, D. Pagliero, A. Laraoui, N.B. Manson, R. Albu, M.W. Doherty, C.A. Meriles, *Nat. Commun.* **7**, 12660 (2016).

[11] See Supplemental Material for further details on the experimental arrangement and analysis methods, which also includes Ref. [12-17].

[12] A.M. Edmonds, U.F.S. D'Haenens-Johansson, R.J. Cruddace, M.E. Newton, K.M.C. Fu, C. Santori, R.G. Beausoleil, D.J. Twitchen, M.L. Markham, *Phys. Rev. B* **86**, 035201 (2012).

[13] L. Rondin, G. Dantelle, A. Slablab, F. Grosshans, F. Treussart, P. Bergonzo, S. Perruchas, T. Gacoin, M. Chaigneau, H. C. Chang, V. Jacques, and J.F. Roch, *Phys. Rev. B* **82**, 115449 (2010).

[14] V. Stepanov, S. Takahashi, *Phys. Rev. B* **94**, 024421 (2016).

[15] L.S. Pan, D.R. Kania, P. Pianetta, O.L. Landen, *Appl. Phys. Lett.* **57**, 623 (1990).

[16] S. Han, L.S. Pan, D.R. Kania, *Diamond: Electronic Properties and Applications*, Eds. L.S. Pan, D.R. Kania (Springer Science, New York, 1995) 241-284.

[17] J. Isberg, M. Gabrysch, J. Hammersberg, S. Majdi, K.K. Kovi, D.J. Twitchen, *Nat. Mater.* **12**, 760 (2013).

[18] R.G. Farrer, *Sol. State Commun* **7**, 685 (1969).

[19] J. Isberg, A. Tajani, D.J. Twitchen, *Phys. Rev. B* **73**, 245207 (2006).

[20] M. Nesládek, L.M. Stals, A. Stesmans, K. Iakoubovskij, G.J Adriaenssens, J. Rosa, M. Vaněček, *Appl. Phys. Lett.* **72**, 3306 (1998).

[21] F.J. Heremans, G.D. Fuchs, C.F. Wang, R. Hanson, D.D. Awschalom, *Appl. Phys. Lett.* **94**, 152102 (2009).

[22] M.C. Rossi, S. Salvatori, F. Galluzzi, *Diam. Rel. Mater.* **6**, 7 12 (1997).

[23] K. Iakoubovskii, G.J. Adriaenssens, *Phys. Rev. B* **61**, 10174 (2000).

[24] J.P. Goss, P.R. Briddon, M.J. Shaw, *Phys. Rev. B* **76**, 075204 (2007).
[25] A. Gali, J.R. Maze, *Phys. Rev. B* **88**, 235205 (2013).
[26] S. Häußler, G. Thiering, A. Dietrich, N. Waasem, T. Teraji, J. Isoya, T. Iwasaki, M. Hatano, F. Jelezko, A. Gali, A. Kubanek, *New. J. Phys.* **19**, 063036 (2017).
[27] A.T. Collins, *J. Phys.: Condens. Matter* **14**, 3743 (2002).
[28] S. Dhomkar, J. Henshaw, H. Jayakumar, C.A. Meriles, *Science Adv.* **2**, e1600911 (2016).
[29] A. Sipahigil, R. E. Evans, D.D. Sukachev, M.J. Burek, J. Borregaard, M.K. Bhaskar, C.T. Nguyen, J.L. Pacheco, H.A. Atikian, C. Meuwly, R.M. Camacho, F. Jelezko, E. Bielejec, H. Park, M. Lončar, M.D. Lukin, *Science* **354**, 847 (2016).
[30] B. L. Green, S. Mottishaw, B. G. Breeze, A. M. Edmonds, U. F. S. D'Haenens-Johansson, M. W. Doherty, S. D. Williams, D. J. Twitchen, and M. E. Newton, Phys. Rev. Lett. **119**, 096402 (2017).